\documentclass[preprint,12pt]{elsarticle}
\usepackage{longtable}
\usepackage{amssymb}
\usepackage{color}
\usepackage{graphicx}
\usepackage{dcolumn}
\usepackage{caption}
\usepackage{amsmath}
\usepackage{booktabs}
\usepackage{multirow}
\usepackage{natbib}

\newcommand{\beq}{\begin{equation}}
\newcommand{\eeq}{\end{equation}}
\newcommand{\beqn}{\begin{eqnarray}}
\newcommand{\eeqn}{\end{eqnarray}}
\journal{Physics Letter B}
\begin{document}

\begin{frontmatter}

\title{Large language model for unified and accurate description of multidimensional nuclear properties}

\author[label1]{S. J. Guo}
\author[label1,label2]{S. Y. Wang\corref{corr1}}\ead{sywang@sdu.edu.cn}
\author[label1,label2]{E. H. Wang}
\author[label3]{Z. M. Niu}
\author[label1]{Y. M. Ding}

\cortext[corr1]{Corresponding author}
\address[label1]{Shandong Provincial Key Laboratory of Nuclear Science, Nuclear Energy Technology and Comprehensive Utilization, Weihai Frontier Innovation Institute of Nuclear Technology, School of Nuclear Science, Energy and Power Engineering, Shandong University, Shandong 250061, People's Republic of China}
\address[label2]{WeiHai Research Institute of Industrial Technology, Shandong University, Weihai 264209, People's Republic of China}
\address[label3]{School of Physics, Anhui University, Hefei 230601, People's Republic of China}

\begin{abstract}
A prior--informed large language model (LLM) driven multi--task learning framework is proposed for the unified description of multiple nuclear observables. By fine--tuning the pre--trained DeepSeek-R1-1.5B model with Low--Rank Adaptation (LoRA), lightweight adapters are introduced while preserving general pre--trained parameters. Under a causal language modeling paradigm, the model is trained autoregressively on deviations between experimental and theoretical values. Significant accuracy improvements are achieved across seven observables, including charge radii, masses, binding energies, separation energies, and decay energies, with the training loss decreasing by over 98\% across all tasks. This demonstrates that the LLM--based framework, through structured prior embedding, offers an efficient and shared approach for multi--task regression in fundamental nuclear properties.
\end{abstract}


\end{frontmatter}

\section{Introduction}

The precise description of the fundamental properties of atomic nuclei has long been one of the central driving forces in nuclear science, engineering and technology research. Nuclear observables not only serve as important labels on the nuclear chart but also provide insight into the complex interactions between nucleons~\cite{J59, K66, P65}. Mass and binding energy characterize the total energy of nuclear system and are essential for determining the stability boundaries and reaction thresholds~\cite{F60, E61}. Single--proton and single--neutron separation energies are sensitive probes of the evolution of shell closures and magic numbers, pairing correlations and weakly bound states in exotic near the drip lines~\cite{P57}. Alpha decay and beta decay are closely related to half--life and nucleosynthesis pathway predictions~\cite{D56, Y58, K55, P54, R64}. Charge radii reflect the spatial properties such as deformation evolution and neutron--skin structure~\cite{A62, M63}. Moreover, charge--radius measurements are relevant to tests of the standard model and to the description of neutron star~\cite{C68, B69}. These seven types of properties collectively form an intricate and inseparable constraint network. In strongly interacting nuclear systems, a unified, self--consistent, and high--precision theoretical description of these properties requires a proper treatment of complex many--body correlations. Achieving an accurate description of these observables remains challenging for existing models~\cite{P57, C51, L52, A53, N41}.

In recent years, data--driven machine learning (ML) methods have opened new gates for the study and prediction of nuclear properties. Tree models~\cite{G15, Z16}, Bayesian analysis~\cite{Z16, J17, Z18, L19, X20, X21}, neural networks~\cite{Z18, K23, S24, X21, Y22} and kernel ridge regression~\cite{X11, X12, J13, X14} have been widely applied to the prediction of nuclear properties. These include decay lifetimes\cite{Z28, A29, N30}, excitation energies~\cite{B32, Z31}, reaction cross sections~\cite{Z33, S34, H35, N36}, charge radii~\cite{Z16, X21, J37, L38, Y39, D40, Y01, T02}, etc. Especially for nuclear masses and single--nucleon separation energies, these models achieve high--precision predictions, with Mean Absolute Error (MAE) and Root Mean Square Error (RMSE) reaching as low as 0.1 MeV to 0.3 MeV~\cite{G15, Z18, X14, Z25, A26, M27}. These advancements fully demonstrate the unique advantages of machine learning in capturing complex nonlinear correlations within nuclear structure data.

High--precision fitting of a single observable does not necessarily imply the self--consistent unification of multi--dimensional physical properties. When the same model framework is applied to the study of multi--dimensional properties~\cite{R10}, the sensitive regions of model parameters for different observations may not mutually compatible. As a result, their loss functions may generate opposing gradient directions in the parameter space, making it difficult for a single model to simultaneously converge to the optimal regions for each objective. To solve this problem, researchers have introduced multi--objective optimization strategies~\cite{Y09}, to find optimal balance solutions in multi--dimensional space. These efforts have produced a series of valuable exploratory results~\cite{R08, X03, X04, W05, C06, J07}.  Nevertheless, existing multi--objective researches still have limitations in predictive accuracy, objective count, feature simplicity, and computational efficiency. Therefore, the optimization of model architectures and the enhancement of model expressiveness have become key issues.

Since the advent of large language models, the transfer value of hundreds of billions or even trillions of pre--trained parameters has increasingly been explored in the scientific domain~\cite{A44, O45, W46, C48}. Compared to traditional ML models, large language models (LLMs), with their vast number of parameters providing substantial model complexity, possess the potential to capture more hidden correlation patterns in nuclear structure data~\cite{X47}. This capability is particularly crucial in multi--objective training processes, where a deeper parameter space can provide a stronger capacity for multi--task learning and a higher precision ceiling. Furthermore, no study has yet been conducted on the fundamental nuclear properties utilizing LLM. Based on this, the present work proposes a LLM--driven multi--objective research framework that integrates prior information. Specifically, we fine--tune a pre--trained LLM (DeepSeek-R1-1.5B)~\cite{D67}, and couple it with physical prior information. This enables the framework to predict a higher--dimensional target outputs by using simpler features, thereby achieving a high--precision unified description of multiple nuclear properties.

\section{Theoretical framework}

\subsection{Input and output definitions}

\begin{figure}[!htb]
\centering
\includegraphics
  [width=0.99\textwidth]
  {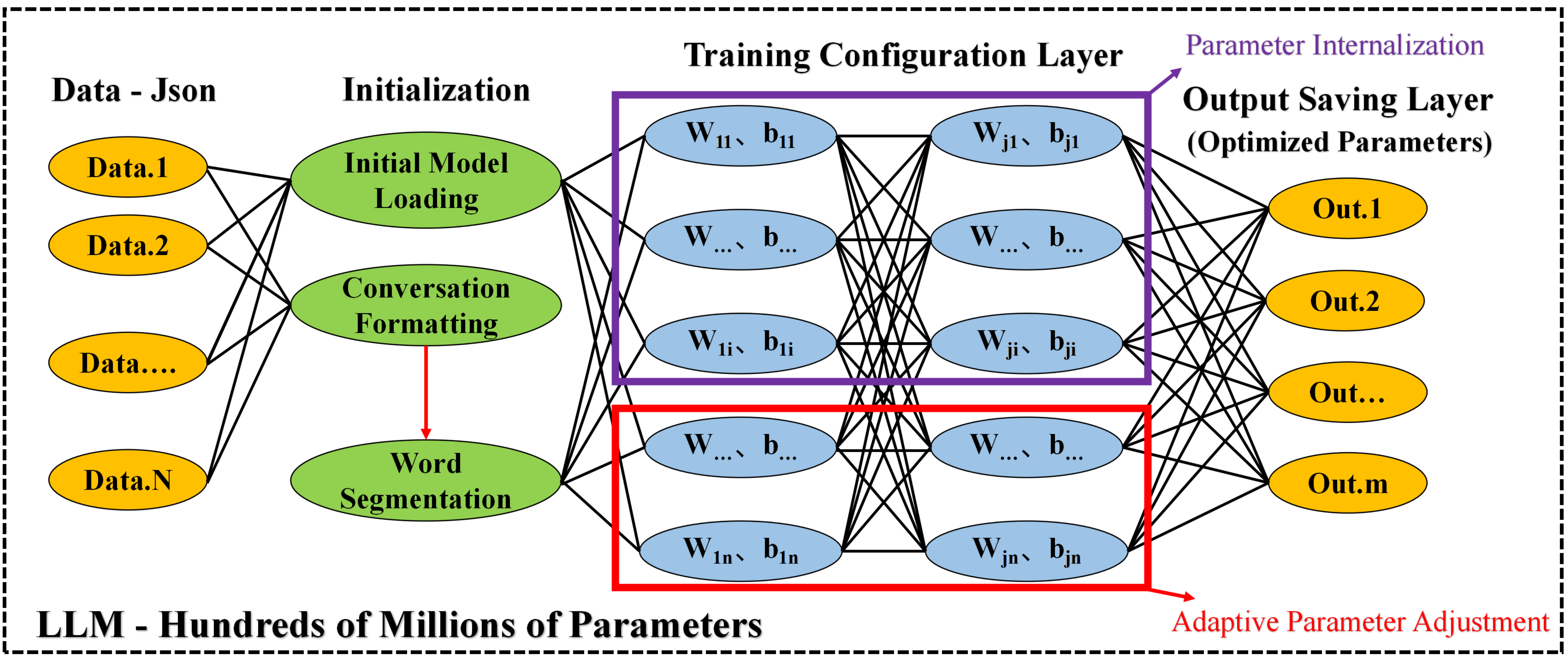}
\caption{(Color online) The large language model architecture.}
\label{fig:The large language model architecture diagram}
\end{figure}

Unlike the traditional regression paradigm that directly models the numerical space, LLM processes data within a generative framework based on natural language sequence generation. Fig.\ref{fig:The large language model architecture diagram} illustrates the logical architecture of the LLM. Taking $N$ and $Z$ as input features, $Q_\alpha$, $Q_\beta$, $mass$, $BE$ as output targets as an example, the original numerical data must first be formatted into a JSON--structured file to adapt to the input sequence parsing specifications of DeepSeek-R1-1.5B:
\begin{equation}
\begin{aligned}
\mathbf{x} &= \mathrm{concat}\big( t_{\mathrm{sys}}, \mathbf{s}, t_{\mathrm{user}}, \mathbf{u}, t_{\mathrm{assist}}, \mathbf{v}, t_{\mathrm{eos}} \big) \\
&= \langle|\mathrm{system}|\rangle + \mathbf{s} + \langle|\mathrm{user}|\rangle + \mathbf{u} + \langle|\mathrm{assistant}|\rangle \\
&+ \mathbf{v} + \langle|\mathrm{endoftext}|\rangle,
\end{aligned}
\end{equation}where $\mathbf{s}$ is system instructions that embed domain--specific prior knowledge, $\mathbf{u}$ encodes the query content including $N$ and $Z$, and $\mathbf{v}$ represents the target response placeholder to be predicted during autoregression. Other special tokens are used to demarcate the structural boundaries of the dialogue. After parsing:
\begin{equation}
\begin{aligned}
\mathbf{x} = &\ \langle|\mathrm{system}|\rangle\ \text{You are ....} \\
&\ \langle|\mathrm{user}|\rangle\ \text{The neutron number is } N\text{,} \\
&\ \text{the proton number is } Z\text{, result is?} \\
&\ \langle|\mathrm{assistant}|\rangle\ \text{The } \alpha\text{-decay energy is } Q_\alpha\text{,} \\
&\ \text{the } \beta\text{-decay energy is } Q_\beta\text{,} \\
&\ \text{the nuclear mass is } mass\text{,} \\
&\ \text{the binding energy is } BE\text{.}\ \langle|\mathrm{endoftext}|\rangle .
\end{aligned}
\end{equation}
The input sequence $\mathbf{x}$ constructed above produces a discrete token series of length $T$, denoted as:
\begin{equation}
\begin{aligned}
\mathbf{x} = [x_1, x_2, \ldots, x_T],
\end{aligned}
\end{equation}where each element $x_t \in \mathcal{V}$ is an integer token identifier from the model's vocabulary, $x_1 = \texttt{<|system|>}$, $x_2$ corresponds to the first token of the system prompt $You$, $x_T = \texttt{<|endoftext|>}$ marks the boundary of the sequence. Under the causal language modeling paradigm, the model is trained to predict the subsequent token based on all preceding context. Correspondingly, the target sequence $\mathbf{y}$ is constructed by shifting the input sequence one position to the right:\begin{equation}
\begin{aligned}
\mathbf{y} = [x_2, x_3, \ldots, x_T].
\end{aligned}
\end{equation}
It contains $T-1$ prediction targets. For any position $t \in {1, 2, \ldots, T-1}$, given the prefix context $[x_1, x_2, \ldots, x_t]$, the model is optimized to maximize the probability of generating the true token $y_t = x_{t+1}$. This autoregressive form enables the model to learn the underlying conditional distribution of token sequences, thereby capturing the syntactic structure of prompt templates and the physical laws embedded within structured data in the kernel.

\subsection{Training process}

The input sequence $\mathbf{x}$ of length $T$ is first mapped to a sequence of high--dimensional embedding vectors via a frozen embedding layer:
\begin{equation}
\begin{aligned}
\mathbf{e}_t = \mathrm{Embedding}(x_t) \in \mathbb{R}^{d}, \quad t = 1, \ldots, T.
\end{aligned}
\end{equation}
Therein, the model dimension $d = 2048$ corresponds to the adopted DeepSeek-R1-1.5B architecture, which is subsequently achieved by stacking $L = 28$ identical transformer layers.

\begin{figure}[!htb]
\centering
\includegraphics
  [width=0.99\textwidth]
  {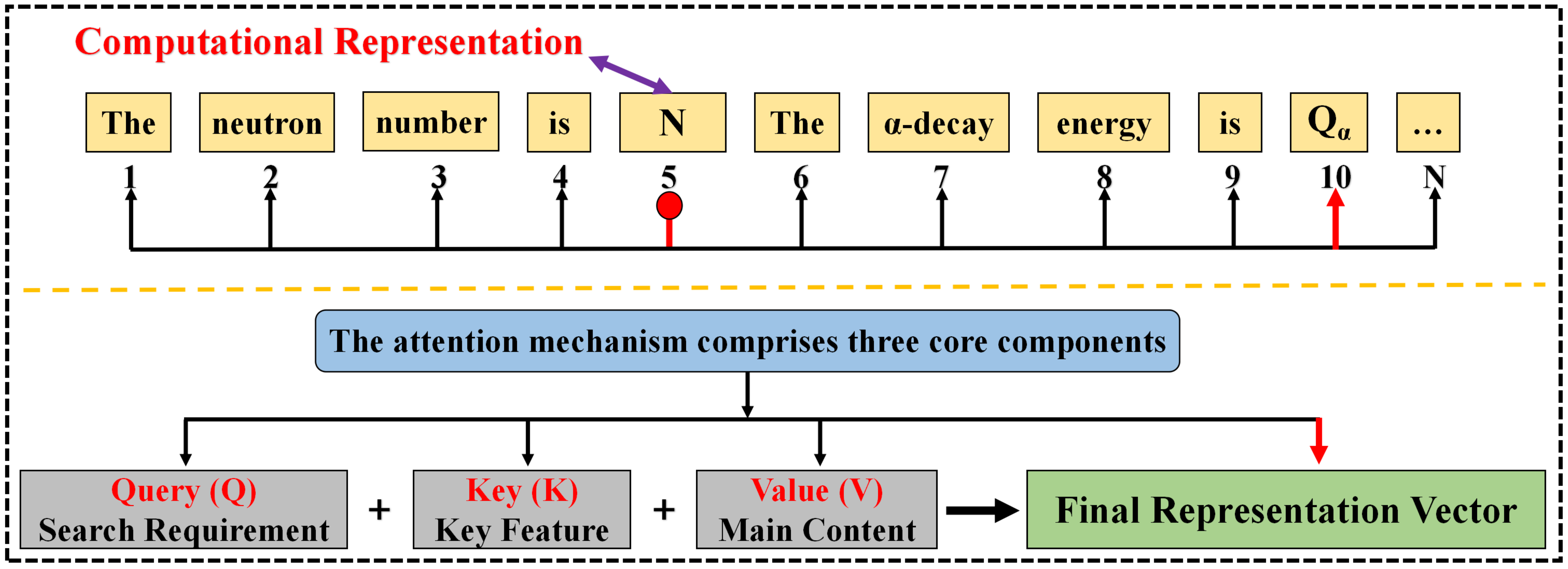}
\caption{(Color online) Schematic diagram of the attention mechanism. When processing the next token, whom will $N$ attend to? Red indicates the highest intensity. In this context, the model attends more strongly to $Q_\alpha$, as $N$ exerts a critical influence on $Q_\alpha$.}
\label{fig:Attention Mechanism}
\end{figure}

Within each layer, the self--attention mechanism (Fig.\ref{fig:Attention Mechanism}) projects the embeddings $\mathbf{e}_t$ into query, key, and value vectors, respectively:
\begin{equation}
\mathbf{q}_t = \mathbf{e}_t \mathbf{W}_q, \quad \mathbf{k}_t = \mathbf{e}_t \mathbf{W}_k, \quad \mathbf{v}_t = \mathbf{e}_t \mathbf{W}_v,
\end{equation}where, $\mathbf{W}_q, \mathbf{W}_k, \mathbf{W}_v \in \mathbb{R}^{d \times d_k}$ are projection matrices. Subsequently, the low--rank adaptation (LoRA) technique~\cite{N49, E50} is employed to efficiently adapt the pre--trained projection matrices to a specific domain.

\begin{figure}[!htb]
\centering
\includegraphics
  [width=0.6\textwidth]
  {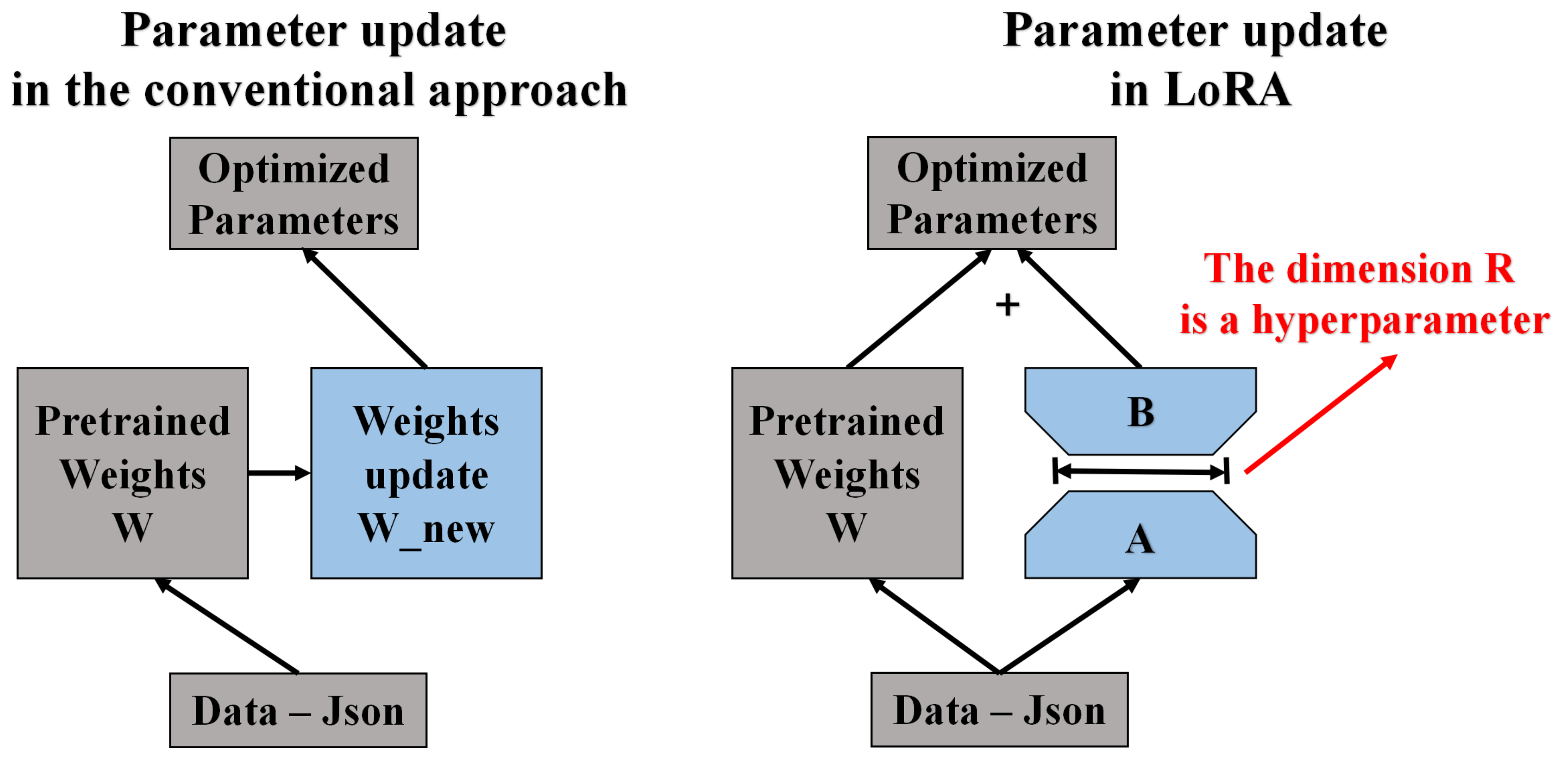}
\caption{(Color online) Comparison of parameter update rules for different approaches.}
\label{fig:Parameter update from LoRA technology}
\end{figure}

Unlike traditional parameter update methods, the update rule of the LoRA technique is illustrated in Fig.\ref{fig:Parameter update from LoRA technology}. The projection matrix is reparameterized as:\begin{equation}
\mathbf{W}_q^{\mathrm{final}} = \mathbf{W}_q^{\mathrm{pretrained}} + \mathbf{B}_q \mathbf{A}_q,
\end{equation}among these, the pre--trained weights $\mathbf{W}_q^{\mathrm{pretrained}}$ are kept frozen, and only the low--rank matrices $\mathbf{B}_q \in \mathbb{R}^{d \times r}$ and $\mathbf{A}_q \in \mathbb{R}^{r \times d_k}$ (rank $r \ll d$) participate in training. For $\mathbf{W}_v$ and the output projection matrix $\mathbf{W}_o$, the same modification approach is adopted.

The attention score~\cite{A44} between position $t$ and position $j$ is computed by the scaled dot--product:
\begin{equation}
\mathrm{score}_{t,j} = \frac{\mathbf{q}_t \cdot \mathbf{k}_j}{\sqrt{d_k}}.
\end{equation}

By computing attention scores, the model can obtain the degree of attention each position pays to other positions. Subsequently, normalization is performed using the softmax function to obtain attention weights, i.e., the attention probability of position $t$ to position $j$:
\begin{equation}
a_{t,j} = \frac{\exp(\mathrm{score}_{t,j})}{\sum_{k=1}^{T} \exp(\mathrm{score}_{t,k})}.
\end{equation}

The context vector after aggregation at position $t$ is given by a weighted sum of the value vectors:
\begin{equation}
\mathbf{z}_t = \sum_{j=1}^{T} a_{t,j} \mathbf{v}_j.
\end{equation}

The output projection $\mathbf{z}_t^{\mathrm{final}}=\mathbf{z}_t \mathbf{W}_o$ is then applied. Residual connections and layer normalization are subsequently applied:
\begin{equation}
\begin{aligned}
&\mathbf{a}_t = \mathrm{LayerNorm}(\mathbf{z}_t^{\mathrm{final}} + \mathbf{e}_t),\\
&\mathrm{LayerNorm}(\mathbf{x}) = \frac{\mathbf{x} - \mu}{\sqrt{\sigma^2 + \epsilon}} \odot \boldsymbol{\gamma} + \boldsymbol{\beta},
\end{aligned}
\end{equation}
$\mu$ and $\sigma^2$ denote the mean and variance, respectively, and $\boldsymbol{\gamma}, \boldsymbol{\beta}$ are learnable scaling and shifting parameters.

The normalized output is further processed by a position--wise feed--forward network:
\begin{equation}
\begin{aligned}
&\mathbf{f}_t = \mathrm{GELU}(\mathbf{a}_t \mathbf{W}_{\mathrm{gate}}) \odot (\mathbf{a}_t \mathbf{W}_{\mathrm{up}}), \\
&\mathbf{h}_t = \mathbf{f}_t \mathbf{W}_{\mathrm{down}},
\end{aligned}
\end{equation}where $\mathrm{GELU}$ is the activation function, and $\odot$ denotes element--wise multiplication. All projection matrices $\mathbf{W}_{\mathrm{gate}}, \mathbf{W}_{\mathrm{up}}$ and $\mathbf{W}_{\mathrm{down}}$ are in a LoRA manner. The second residual connection and layer normalization yield the final output of the layer:
\begin{equation}
\mathbf{h}_t^{\mathrm{final}} = \mathrm{LayerNorm}(\mathbf{h}_t + \mathbf{a}_t).
\end{equation}

After propagation through all $L$ layers, the resulting hidden layer $\mathbf{h}_t^{\mathrm{final}, L}$ is passed to the output layer. The final hidden state at each position is mapped to a probability distribution over the vocabulary $\mathcal{V}$ via an output projection matrix $\mathbf{W}_{\mathrm{head}}$:
\begin{equation}
\begin{aligned}
\mathbf{o}_t &= \mathbf{h}_t^{\mathrm{final},L} \mathbf{W}_{\mathrm{head}} \in \mathbb{R}^{|\mathcal{V}|}, \\
\mathbf{p}_t &= \mathrm{softmax}(\mathbf{o}_t), \\
p_t(k) &= \frac{\exp(o_{t,k})}{\sum_{j=1}^{|\mathcal{V}|} \exp(o_{t,j})}.
\end{aligned}
\end{equation}

The loss function for a single sequence is defined as the average negative log--likelihood over $T-1$ predicted positions:
\begin{equation}
\mathcal{L}(\boldsymbol{\Theta}) = -\frac{1}{T-1} \sum_{t=1}^{T-1} \log p_t(y_t),
\end{equation}where $p_t(y_t)$ is the predicted probability for the ground truth label $y_t$. The total loss on the training set $\mathcal{D}$ is the empirical average:
\begin{equation}
\mathcal{L}_{\mathrm{total}}(\boldsymbol{\Theta}) = \frac{1}{|\mathcal{D}|} \sum_{k=1}^{|\mathcal{D}|} \mathcal{L}_k(\boldsymbol{\Theta}).
\end{equation}

Trainable parameters $\boldsymbol{\Theta}$ only contain LoRA matrices $\{\mathbf{B}_q, \mathbf{A}_q, \mathbf{B}_v, \mathbf{A}_v, \ldots\}$ in all layers. These parameters are determined by minimizing a loss function via stochastic gradient descent. The gradients of the loss with respect to the low--rank matrices (e.g., $\mathbf{A}_q$) are computed via backpropagation using the chain rule, and the parameters are updated iteratively according to the following formula:
\begin{equation}
\mathbf{A}_q^{\mathrm{new}} = \mathbf{A}_q^{\mathrm{old}} - \eta \frac{\partial \mathcal{L}}{\partial \mathbf{A}_q}.
\end{equation}

Here, $\eta$ denotes the learning rate. All weight matrices adapted by LoRA follow the same update rule. The essence of this method lies in introducing lightweight adapters trained on a specific domain, while preserving the frozen pre--trained weights for general language and reasoning capabilities. By reconstructing and learning multiple related target quantities, the model can leverage the deeper implicit shared structural information within the data, thereby enhancing the generalization ability for all targets.

\section{Data preparation and model design}

Data sources~\cite{I43, N41, M42} and statistics of preprocessed data are shown in Tab.\ref{tab:distA}. The residuals between experimental values and theoretical values are taken as the training target by the model. In the theoretical model construction, $Q_\alpha$ was predicted using the Random Forest algorithm, while $Q_\beta$ was predicted based on the LightGBM algorithm. The evaluation results on the testing set indicate that the MAE and RMSE of $\mathrm{ML}_{\mathrm{RF}}$ are 0.2259 MeV and 0.3852 MeV, respectively, while the corresponding errors for $\mathrm{ML}_{\mathrm{LightGBM}}$ are 0.1447 MeV and 0.3754 MeV. Additionally, $r$ is given by a semi--empirical formula, and the other four parameters are derived from the mass model WS4.

\begin{figure}[!htb]
\centering
\includegraphics
  [width=0.8\textwidth]
  {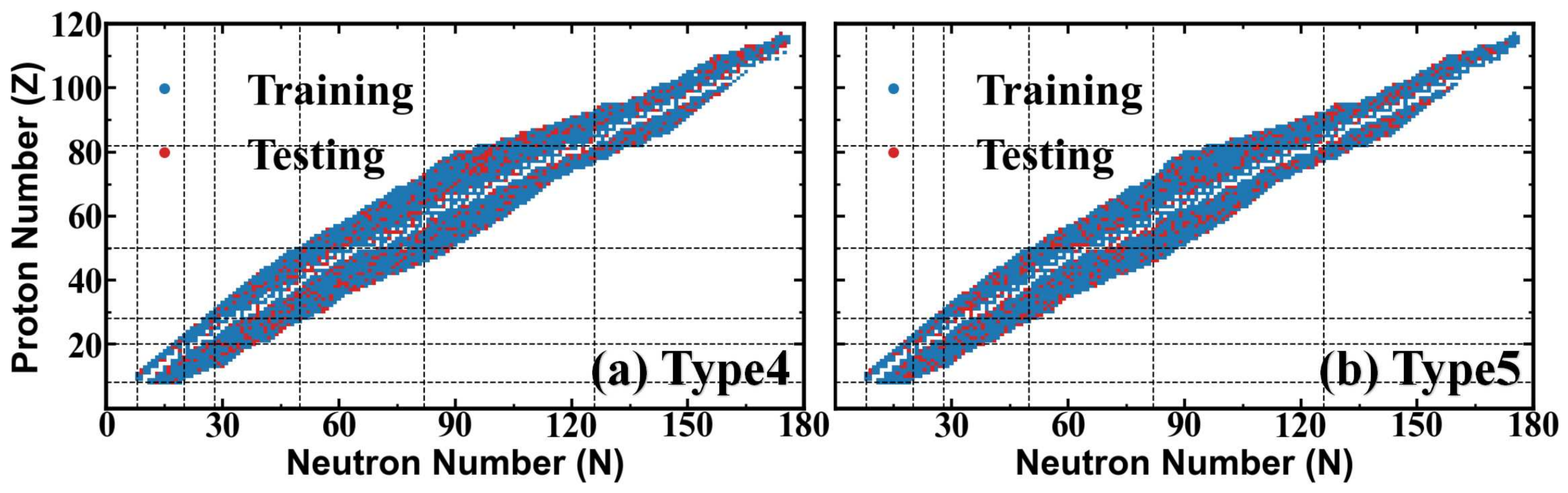}
\caption{(Color online) The data distributions of the training and testing sets in Type4 (a) and Type5 (b). The dashed lines in the figure indicate the positions of magic numbers.}
\label{fig:traintest}
\end{figure}

To achieve multi--dimensional comparative analysis, five distinct training paradigms were constructed, with the feature combinations for each paradigm defined as shown in Tab.\ref{tab:distB}. Across all task categories, the dataset was divided into training and testing sets with a ratio of 8:2. The data distributions of the training and testing sets for Type4 and Type5 are shown in Fig.\ref{fig:traintest}. The testing set was not involved in model training at any stage and was solely utilized for evaluating generalization performance post--training, thereby preventing data leakage from compromising result fidelity. Two evaluation metrics, MAE and RMSE, were employed to quantify the performance of the LLM for comparison with other reference results.

\section{Results and discussion}

\subsection{Single--task prediction of nuclear charge radii}

\begin{figure}[!htb]
\centering
\includegraphics
  [width=0.8\textwidth]
  {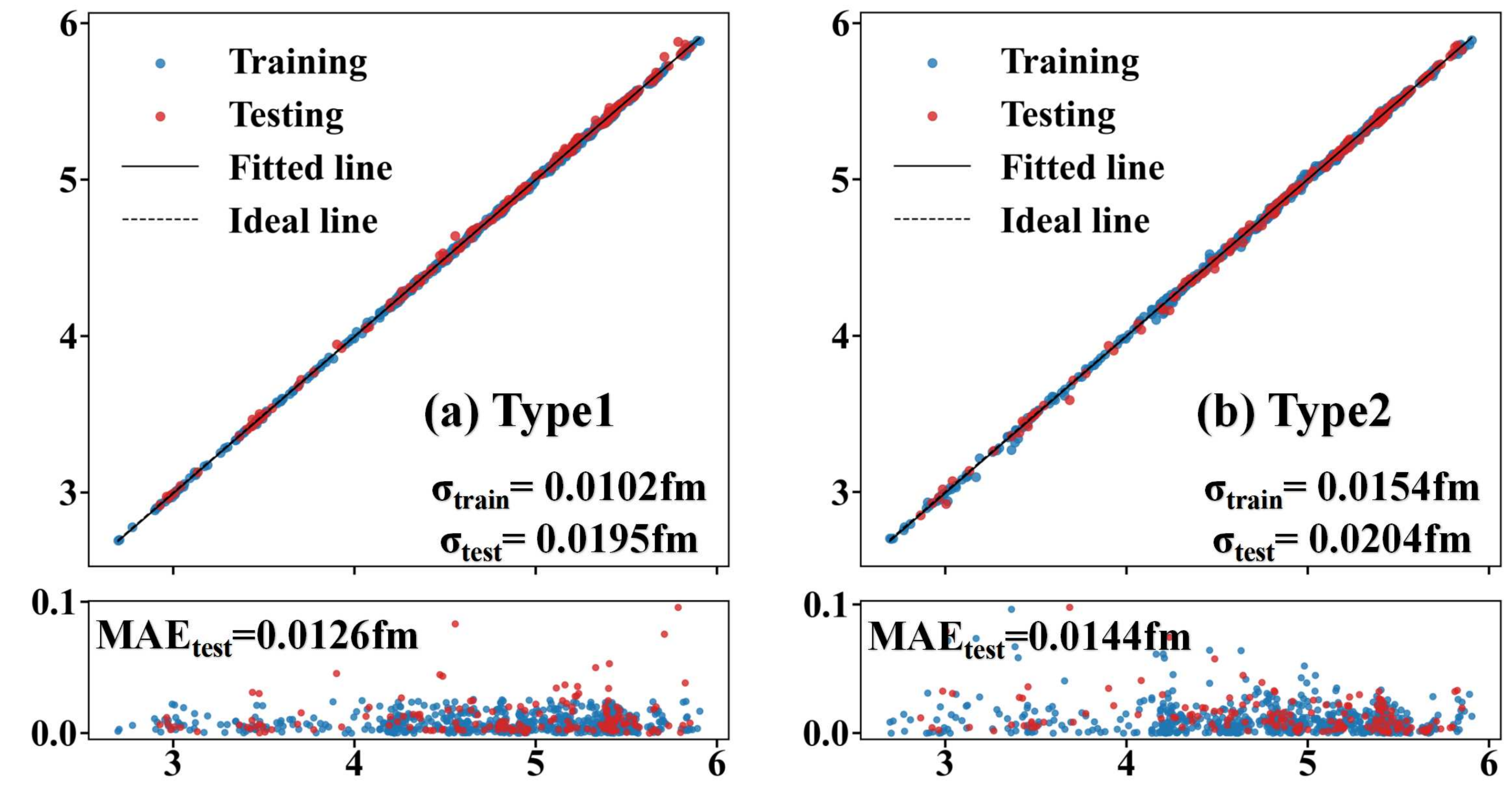}
\caption{(Color online) Comparison between experimental and predicted $r$ in the region $Z \geq 8$. (a) Results from Type 1 with five input features, showing the fitting curve and outlier distribution. (b) Results from Type 2 with only $N$ and $Z$ as input features. The model accurately captures the main trend of charge radii even with minimal input, though precision degrades slightly in the light--nucleus region compared to Type 1.}
\label{fig:r}
\end{figure}

To reveal the characteristics of the LLM in processing regression tasks, a single--task learning framework is first employed to predict the charge radius. Type1 takes $N$, $Z$, $M_p$, $M_n$ and $\delta$ as input features, where $M_p$ ($M_n$) denotes the distance between the proton (neutron) number and the nearest magic number, and $\delta = \frac{1}{2}[(-1)^Z + (-1)^N]$ for characterizing the nucleon pairing shell effect. After training, across the entire dataset, the model achieved MAE and RMSE of 0.0086 fm and 0.0167 fm on the training set, and 0.0136 fm and 0.0221 fm on the testing set, respectively. In the $Z \geq 8$ region, the RMSE for the training and testing sets further decreased to 0.0102 fm and 0.0195 fm, exhibiting a slight accuracy advantage compared with Ref.~\cite{Y01, T02}. The fitting curve and outlier distribution are shown in Fig.\ref{fig:r}(a).

Unlike traditional ML, which relies on manually constructed features, the LLM theoretically possesses the capability to automatically extract effective representations from raw inputs. In this regard, Type2 analyzes $r$ using only $N$ and $Z$ as input. Under this approach, the model achieves MAE and RMSE of 0.0139 fm and 0.0467 fm on the training set for all regions, and 0.0198 fm and 0.0515 fm on the testing set. Compared to Type1, the RMSE increase exceeds a factor of 2. In the region $Z \geq 8$, the dual--feature scheme achieves RMSE values of 0.0154 fm and 0.0204 fm on the training and testing sets, respectively, which are at almost the same level as Ref.~\cite{Y01, T02} and Type1, effectively demonstrating the applicability of the LLM for regression tasks. Fig.\ref{fig:r}(b) also illustrates the fitting curves and outlier distribution.

Using only $N$ and $Z$, the LLM can capture the main trend of $r$ variation. This stems from the autonomous feature extraction capability brought by the deeper and more complex architectural advantages of the LLM: even for simple input features, the model can autonomously extract the underlying deep correlations. However, for both Type1 and Type2, the accuracy in the $Z \geq 8$ region is higher than that in the entire region, revealing that the light nuclear region, due to the complexity of its few--body correlation and cluster structure, has always been a more challenging domain for charge radius prediction. This phenomenon also indirectly indicates the necessity of more prior information: the prior information extractable from $N$ and $Z$ alone has inherent limitations in the light nuclear region. When physical features, including magic numbers and shell effects, are explicitly introduced, the model's prediction error can be effectively suppressed.

\begin{figure}[!htb]
\centering
\includegraphics
  [width=0.5\textwidth]
  {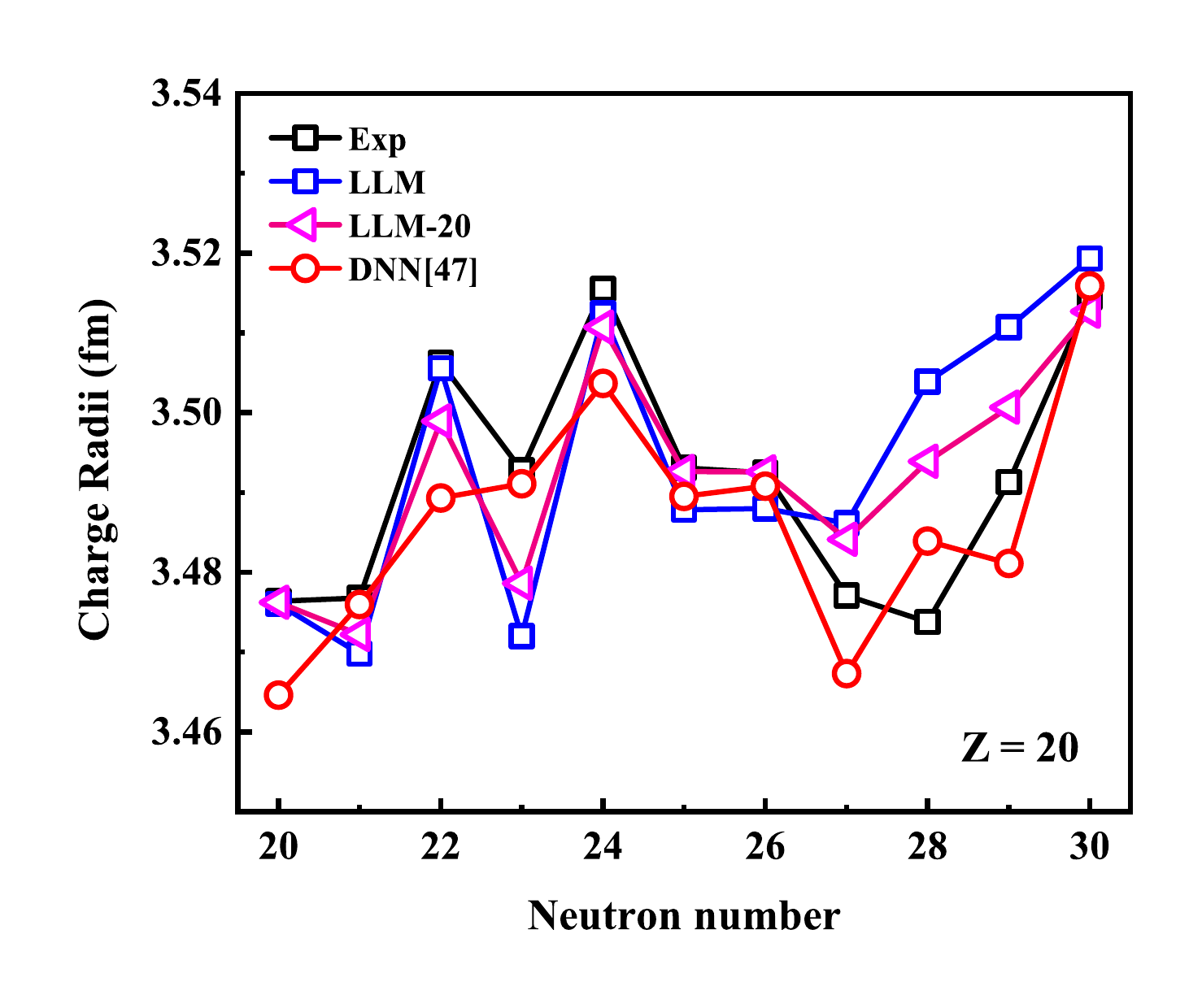}
\caption{(Color online) Comparison of experimental charge radii with predictions from LLM (based on the trained Type1), LLM--20 (mean of 20 predictions), and BNN from Ref.~\cite{Y01} for $Z=20$ with $N$ ranging from 20 to 30.}
\label{fig:x}
\end{figure}

To evaluate the framework's stability and extrapolation capability, the trained model is used for 20 predictions. The predicted values fluctuated by no more than 0.7\%. Fig.\ref{fig:x} shows the comparison results for $Z=20$ with $N$ ranging from 20 to 30.

\subsection{Multi--task prediction}

Given that simple input features did not significantly negatively impact accuracy in the single--objective task, $N$ and $Z$ are directly used for training in the multi--objective task: Type3 simultaneously predicted $mass$, $BE$, $S_p$ and $S_n$. The MAE values on the training and testing sets were between 0.20 MeV and 0.30 MeV, and the RMSE was between 0.30 MeV and 0.55 MeV. The numerical accuracy was slightly insufficient, but it is noteworthy that the prediction accuracy of $mass$ and $BE$ improved almost synchronously, demonstrating that the model can effectively capture the implicit equality constraint between them. When the output targets were replaced from $mass$ and $BE$ to $Q_\alpha$ and $Q_\beta$, the RMSE remained between 0.25 MeV and 0.45 MeV, showing no significant improvement. This indicates that the presence of implicit equality constraints between target tasks is not the dominant factor determining the accuracy level.

\begin{figure*}[!htb]
\centering
\includegraphics
  [width=0.99\textwidth]
  {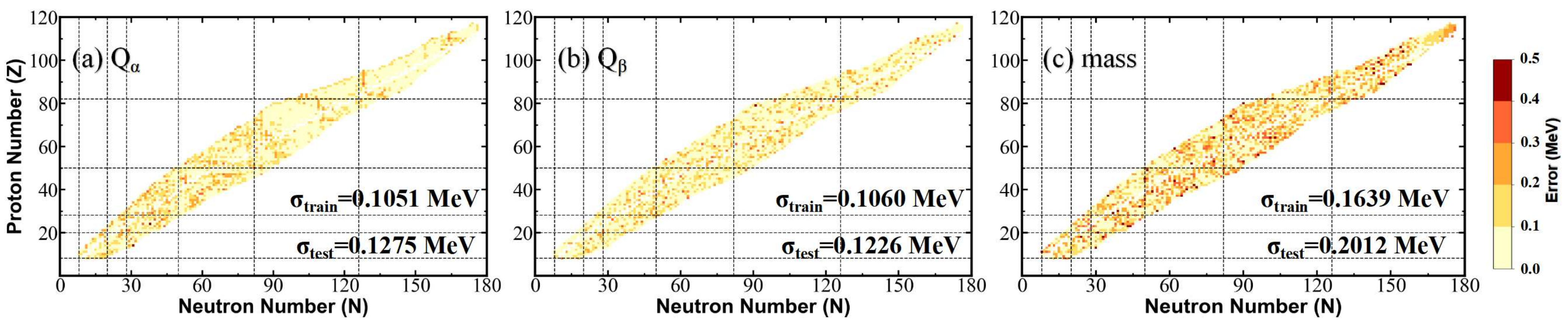}
\caption{(Color online) Predicted versus experimental values for multi--task learning with Type 4 (input: $N$ and $Z$). The deviations are shown for all four observables. The RMSE for $Q_\alpha$ and $Q_\beta$ on the testing set is below 130 keV, while those for $mass$ and $BE$ are around 200 keV. The dashed lines in the figure indicate the positions of magic numbers.}
\label{fig:1}
\end{figure*}

In contrast, Type4 targets $mass$, $BE$, $Q_\alpha$, $Q_\beta$, with the precision of each physical quantity presented in Tab.\ref{tab:distC}. The MAE for $Q_\alpha$ and $Q_\beta$ in the testing set reached approximately 0.09 MeV, with RMSE within 0.13 MeV, while $mass$ corresponded to 0.16 MeV and 0.20 MeV, respectively, all exhibiting consistent performance improvements~\cite{X03, X04, W05}. The errors between predicted and true values for all data are plotted in Fig.\ref{fig:1}. The model's systematic underestimation for outputs $S_p$ and $S_n$ stems from the data structure of these two physical quantities and nuclear many--body effects. The definition of nucleon separation energy is as follows:
\begin{equation}
\begin{aligned}
S_n(N, Z) &= B(N, Z) - B(N-1, Z), \\
S_p(N, Z) &= B(N, Z) - B(N, Z-1).
\end{aligned}
\end{equation}

The separation energy is essentially a differential operation on the binding energies of adjacent nuclides. This structure amplifies the mutation of the binding energy surface near magic numbers, leading to significant local fluctuations and non--smooth features in the data distribution. When only $N$ and $Z$ are used as input, the model does not possess any explicit shell prior information, making direct learning difficult. Conversely, $Q_\alpha$ and $Q_\beta$, during their construction, systematically cancel out a large number of local fluctuations influenced by shell effects, making the overall function smoother and thus easier for the model to fit. This explains why, under the same input conditions, the prediction error for separation energy is higher than that for decay energy, and further confirms that the data smoothness of the target physical quantity itself plays a decisive role during training.

\begin{figure*}[!htb]
\centering
\includegraphics
  [width=0.99\textwidth]
  {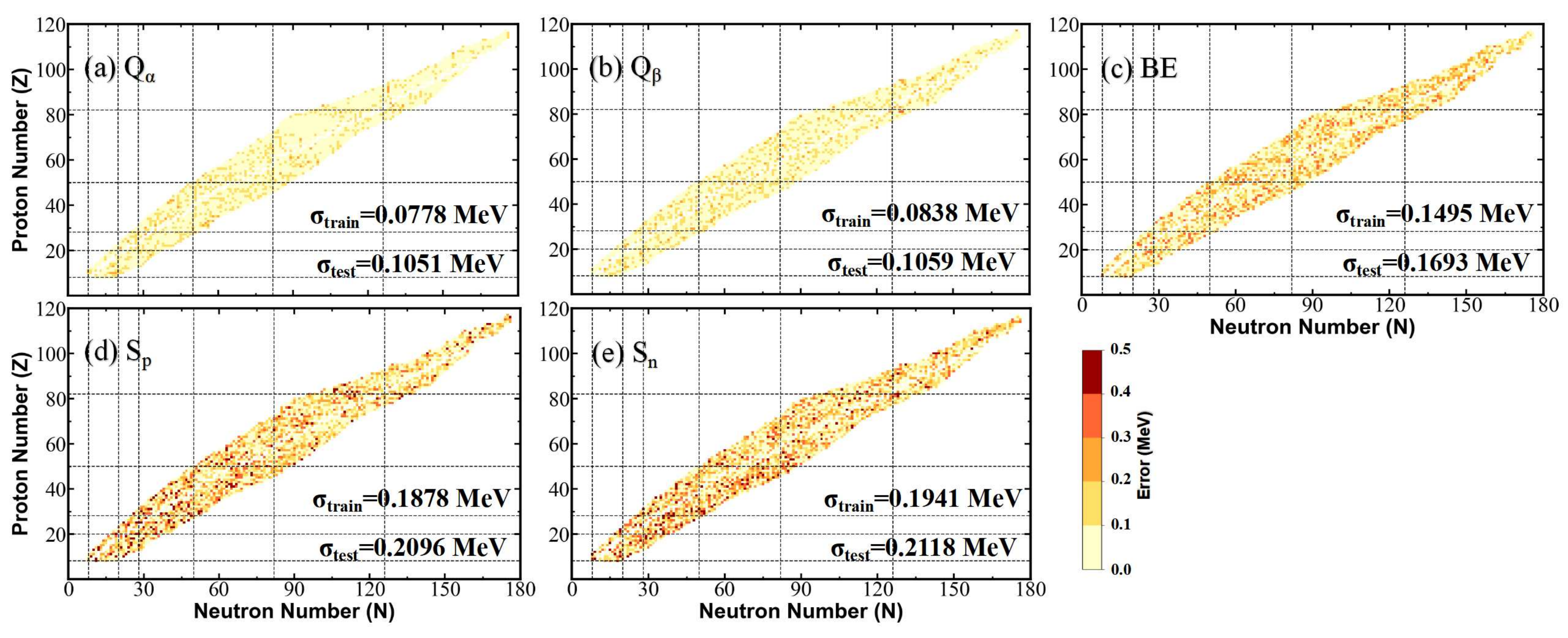}
\caption{(Color online) Predicted versus experimental values for multi--task learning with Type 5. The incorporation of physical quantities related to the magic numbers and shell effects further reduces the prediction errors, with RMSE around 100 keV for $Q_\alpha$ and $Q_\beta$, below 170 keV for $BE$, and approximately 200 keV for $S_p$ and $S_n$. The dashed lines in the figure indicate the positions of magic numbers.}
\label{fig:2}
\end{figure*}

To compensate for this deficiency and maximize the prediction accuracy of various physical quantities as much as possible, physical quantities related to magic numbers and shell effects, $M_p$, $M_n$ and $\delta$ are explicitly introduced as additional input features. The performance of the trained Type5 on the results confirms the effectiveness of this approach, with detailed results listed in Tab.\ref{tab:distC}. On the training set, the MAE for $Q_\alpha$ and $Q_\beta$ reached within 0.065 MeV, and the RMSE was around 0.080 MeV. For $S_p$ and $S_n$, the prediction accuracy was also significantly improved, decreasing to within 0.150 MeV and 0.195 MeV, respectively, with BE reaching 0.116 MeV and 0.149 MeV. On the testing set, the errors for each physical quantity were highly consistent with the training set, with deviations controlled within 0.020 MeV. Specifically, the MAE for $Q_\alpha$ and $Q_\beta$ was around 0.080 MeV, and the RMSE was around 0.105 MeV. $S_p$ and $S_n$ corresponded to 0.170 MeV and 0.210 MeV, respectively, with BE reaching 0.135 MeV and 0.169 MeV~\cite{X03, X04, W05}. Fig.\ref{fig:2} shows the errors between the predicted and true values for each physical quantity. The good consistency between the training and testing sets indicates the reliability of its generalization ability.

\begin{figure}[!htb]
\centering
\includegraphics
  [width=0.5\textwidth]
  {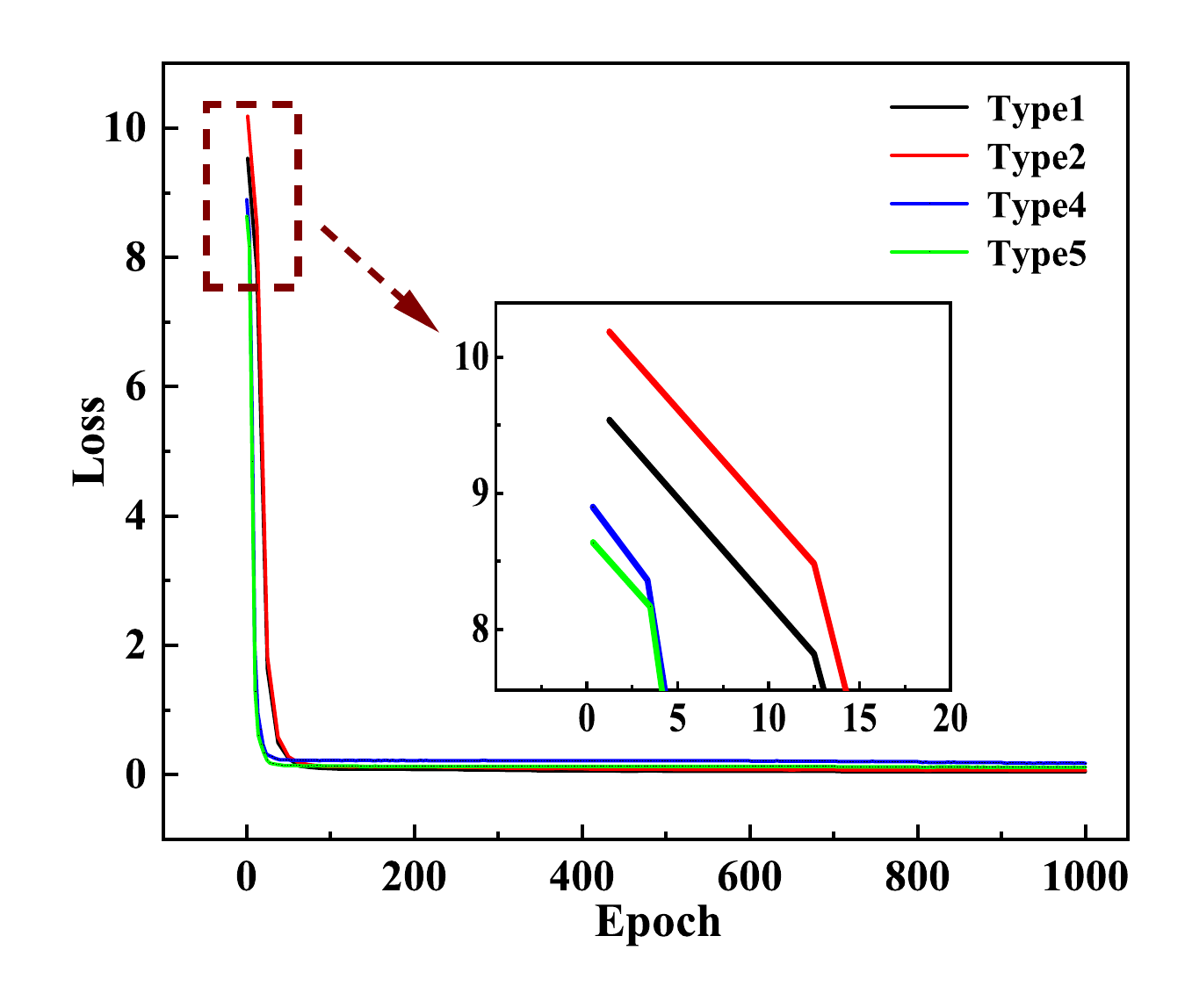}
\caption{(Color online) Convergence trends of the training loss for different task configurations.}
\label{fig:Loss convergence trend}
\end{figure}

The loss variation trends and proportions for various tasks are summarized in Fig.\ref{fig:Loss convergence trend} and Tab.\ref{tab:distD}. During the training process, the loss value decreased significantly: the single--objective task decreased by over 99\%, and the optimization range for the multi--objective task remained stable above 98\%. When the input features contain more prior physical information (magic numbers, shell effects), the model exhibits lower initial loss and a higher optimization proportion. This also validates the positive role of prior information in the LLM framework, but its improvement is more limited compared to ML. This difference once again demonstrates the inherent advantages of the complex architecture of the LLM compared to traditional ML approaches.

\section{Summary}

In summary, a large language model--driven multi--objective research framework integrating prior information is proposed for the study of nuclear observables. By adopting LoRA technology, this framework uses lightweight adapters tailored for specific nuclear property tasks while preserving the general pre--trained parameters. Under the causal language modeling paradigm, the model undergoes autoregressive training on the residuals between experimental data and theoretical models.

In the single--objective task, predicting $r$ based on five input features, the MAE and RMSE on the testing set in the region where $Z \geq 8$ were 0.0126 fm and 0.0195 fm, respectively. In the multi--objective task, using only $N$ and $Z$ as input features to train $Q_\alpha$, $Q_\beta$, $mass$ and $BE$, the model achieved RMSE below 130 keV for $Q_\alpha$ and $Q_\beta$ on the testing set, around 200 keV for $mass$ and $BE$, and effectively captured the intrinsic correlation between $mass$ and $BE$. When the input features were expanded to five, and the output features were $Q_\alpha$, $Q_\beta$, $BE$, $S_p$ and $S_n$, the RMSE for $Q_\alpha$ and $Q_\beta$ in the testing set reached approximately 100 keV, $BE$ improved to within 170 keV, while $S_p$ and $S_n$ were around 200 keV. Compared to traditional ML methods, various tasks demonstrated consistent performance improvements. Meanwhile, during the training process, the loss value showed a significant downward trend: the decrease exceeded 99\% in the single--objective task and remained stable above 98\% in the multi--objective task, validating the robustness and effectiveness of the framework.

The aforementioned results indicate that by incorporating structured prior embeddings, the LLM--based framework establishes a novel and efficient approach for multi--objective regression tasks in fundamental nuclear properties.

Although LLMs can achieve decent accuracy across various tasks, they also have certain limitations. More complex architectures, while improving performance, tend to enhance model interpretability challenges. Meanwhile, the high computational cost restricts their application scenarios. This is one of the direction that requires discussion and exploration.

\section*{Acknowledgments}

The numerical calculations in this work have been done on the super computing system in Shandong University, Weihai. This work is partly supported by the National Natural Science Foundation of China (Grants No. 12225504).

\clearpage

\clearpage
\begin{table}[h]
\centering
\scriptsize
\caption{Data description.}
\label{tab:distA}
\begin{tabular}{ccccc}
\toprule
Feature name & Description & Experimental value & Theoretical value & Dataset size \\
\midrule
$r$      & nuclear charge radius & \cite{I43} & $1.1 \times A^{1/3}$   & 799  \\
$mass$   & nuclear mass         & \multirow{4}{*}{\centering \cite{N41}} & \multirow{4}{*}{\centering WS4} & \multirow{4}{*}{\centering 3231} \\
$BE$     & binding energy        &  &  &  \\
$S_p$    & single-proton separation energy & & & \\
$S_n$    & single-neutron separation energy & & & \\
$Q_\alpha$ & $\alpha$--decay energy   & \multirow{2}{*}{\centering \cite{M42}} & $\mathrm{ML}_{\mathrm{RF}}$ & 3413 \\
$Q_\beta$  & $\beta$--decay energy   & & $\mathrm{ML}_{\mathrm{LightGBM}}$ & 3166 \\
\bottomrule
\end{tabular}
\end{table}

\begin{table}[h]
\centering
\scriptsize
\caption{Model design.}
\label{tab:distB}
\begin{tabular}{cccc}
\toprule
Paradigm & Input features & Output features & Size \\
\midrule
Type1 & $N$, $Z$, $M_p$, $M_n$, $\delta$ & $r$ & 799 \\
Type2 & $N$, $Z$ & $r$ & 799 \\
Type3 & $N$, $Z$ & $mass$, $BE$, $S_p$, $S_n$ & 3231 \\
Type4 & $N$, $Z$ & $mass$, $BE$, $Q_\alpha$, $Q_\beta$ & 2986 \\
Type5 & $N$, $Z$, $M_p$, $M_n$, $\delta$ & $Q_\alpha$, $Q_\beta$, BE, $S_p$, $S_n$ & 2864 \\
\bottomrule
\end{tabular}
\end{table}

\begin{table}[h]
\centering
\scriptsize
\caption{The RMSE(MeV) for Each Multi--task Objective.}
\label{tab:distC}
\begin{tabular}{cccccccc}
\toprule
Data & Task & $Q_\alpha$ & $Q_\beta$ & $mass$ & $BE$ & $S_p$ & $S_n$ \\
\midrule
\multirow{2}{*}{Training} & Type4 & 0.1051 & 0.1060 & 0.1639 & 0.1638 & - & - \\
 & Type5 & 0.0778 & 0.0838 & - & 0.1495 & 0.1878 & 0.1941 \\
\multirow{2}{*}{Testing} & Type4 & 0.1275 & 0.1226 & 0.2012 & 0.2011 & - & - \\
 & Type5 & 0.1051 & 0.1059 & - & 0.1693 & 0.2096 & 0.2118 \\
\bottomrule
\end{tabular}
\end{table}

\begin{table}[h]
\centering
\scriptsize
\caption{Loss variation values across tasks.}
\label{tab:distD}
\begin{tabular}{cccc}
\toprule
Task & Initial loss & Final loss & Improvement \\
\midrule
Type1 & 9.5360 & 0.0459 & 99.52\% \\
Type2 & 10.1834 & 0.0682 & 99.33\% \\
Type4 & 8.8967 & 0.1775 & 98.13\% \\
Type5 & 8.6367 & 0.1105 & 98.72\% \\
\bottomrule
\end{tabular}
\end{table}

\end{document}